\documentclass[12pt]{article}
\usepackage{epsfig, amssymb}
\usepackage{amsmath}
\usepackage{graphicx,epsfig}
\usepackage{color} 
\usepackage{cite}
\begin{document}
\thispagestyle{empty}
\begin{flushright}
\end{flushright}

\bigskip

\begin{center}
\noindent{\Large \textbf
{On holographic Wilsonian renormalization group of massive scalar theory with its self-interactions in AdS
}}\\ 
\vspace{2cm} \noindent{Gitae Kim${}^{a}$\footnote{e-mail:kimgitae728@gmail.com} and 
Jae-Hyuk Oh${}^{a}$\footnote{e-mail:jaehyukoh@hanyang.ac.kr}}

\vspace{1cm}
  {\it
Department of Physics, Hanyang University, Seoul 04763, Korea${}^{a}$\\
 }
\end{center}

\vspace{0.3cm}
\begin{abstract}
\noindent
Holographic model of massive scalar field with its self-interaction $\lambda\phi^n$ in AdS space is able to give a logarithmic scale dependence to marginal multi trace deformation couplings on its dual conformal field theory, where $\lambda$ is the self-interaction coupling of the scalar field, $\phi$ and $n$ is an integral number. In arXiv:1501.06664, the authors realize this feature by looking at bulk scalar solutions near AdS boundary imposing a specific boundary condition between the coefficients of non-normalizable and normalizable modes of the scalar field excitations.
We study the same holographic model to see scale dependence of marginal deformations on the dual conformal field theory by employing completely different method: {\it holographic Wilsonian renormalization group}. We solve Hamilton-Jacobi equation derived from the holographic model of massive scalar with $\lambda\phi^n$ interaction and obtain the solution of marginal multi trace deformations up to the leading order in $\lambda$. It turns out that the solution of marginal multi trace deformation also presents logarithmic behavior in energy scale near UV region.
\end{abstract}


\newpage

\section{Introduction}
Gauge/gravity daulity(holography) has shed lights on Large $N$ gauge field theories.
Especially, the duality proposes precise map between $d$-dimensional conformal field theories and $d+1$-dimensional gravity theories defined in AdS(Anti de Sitter) space\cite{Aharony:1999ti,Maldacena:1997re}. The duality frame work is useful if the $d+1$-dimensional gravity theory is weakly coupled, since it is well known that the weak gravity can probe $d$-dimensional strongly coupled field theory defined on the conformal boundary of AdS space.

In the weak gravity limit, one can formulate a holographic model in $d+1$ dimensional space(especially if one considers a dual conformal field theory, it would be $d+1$ dimensional AdS space) to probe a certain $d$-dimensional field theory.
One may consider a weak gravity action as
\begin{equation}
S=\int_{r \geq 0}d^{d+1}x \left(\frac{1}{2}\partial_\mu \phi \partial^\mu \phi +V(\phi)\right)+ \int_{r=0}d^dxW(\phi,\partial \phi),
\end{equation}
where the $r$ is AdS radial variable, the coordinate variable $x^\mu=(r,x^1,...,x^d)\equiv(r,x^i)$ and $r=0$ is the conformal boundary of AdS space. The usual map of holography is that once one turns on the bulk modes, the modes affect the boundary field theory in the two different ways. The non-normalizable modes of the excitations add certain deformations(or turning on a certain operator) and the normalizable modes of the excitations turn on non-trivial states in the dual field theory\cite{Balasubramanian:1998sn,Balasubramanian:1998de}. The modes satisfy the bulk equations of motion and the boundary conditions respectively as
\begin{equation}
\label{eom-bc}
\nabla^2 \phi+\frac{\delta V(\phi)}{\delta \phi}=0, {\ \ \rm and \ \ }\left.-\frac{\partial \phi}{\partial r}+\frac{\delta W(\phi,\partial \phi)}{\delta \phi}\right|_{r=0}=0,
\end{equation}
in the classical gravity limit, which is given by variation of the action $S$, i.e. $\delta S=0$. The formal form of the bulk excitations is given as
\begin{equation}
\phi(r,x^\mu)=\beta(x^\mu)r^{d-\Delta}+\alpha(x^\mu)r^\Delta,
\end{equation}
where the $\beta$ is the coefficient of the non-normalizable mode, the $\alpha$ is the coefficient of the normalizable mode and $\Delta$ is the conformal dimension of a certain operator turned on the dual field theory. 

The particle states in the gravity theory correspond to turning on two different types of operators in the dual gauge field theory. Once we turn on a single particle state, it corresponds to single trace operator while if we turn on multi particle  states, then it maps to multi trace operator. The multi particle states have been widely studied to answer the following interesting question. Precisely what kind of conditions on the multi particle states will give a specific type of multi trace deformation on the dual gauge field theory? 

The answer is suggested in \cite{Witten:2001ua,Aharony:2015afa}.
The authors in these papers
argue that the boundary condition given in (\ref{eom-bc}) would provide a certain relation between the coefficients of the non-normalizable and normalizable modes of the solutions, and it gives a multi-trace deformation. More precisely, if one gets a boundary condition as
\begin{equation}
\label{multitrace-relation}
\beta(x^\mu)\sim D^{(n)} \alpha^{n-1}(x^\mu),
\end{equation}
then, it means that the bulk non-normalizable modes deform the boundary field theory with multi trace operator with a form of $\int d^dx D^{(n)} O^n(x^\mu)$, where the $O(x^\mu)$ is a certain single trace operator and $D^{(n)}$ is the multi trace coupling. 

This suggestion provides a convincing example in the case that the deformation in the dual conformal field theory is classically marginal. For instance, let us consider a double trace operator deformation on a certain conformal field theory in 4-dimensional spacetime. The deformation is $\frac{D^{(2)}}{2}\int d^4x O^2(x)$, where we set the scaling dimension of the operator $\Delta[O(x)]=2$ for it to be classically marginal and the $D^{(2)}$ is the double trace coupling. The 2-point correlation function of the operator is restricted by conformal symmetry as
\begin{equation}
\langle O(x)O(y) \rangle=\frac{a}{|x-y|^4},
\end{equation}
where the constant, $a>0$ for unitarity. Once such a deformation exists, the conformal field theory partition function changes as
\begin{equation}
\label{holo-partition}
\mathcal Z=\langle  e^{-\frac{D^{(2)}}{2}\int d^4x O^2(x)}\rangle_{\rm CFT}  \rightarrow 
e^{-\frac{D^{(2)}_{\rm eff}}{2} \int d^4x \langle O^2(x) \rangle+ {\rm order\ of \ }\langle O^4(x)\rangle},
\end{equation}
where the right hand side of the arrow is its effective action, with the effective multi trace coupling $D^{(2)}_{\rm eff}$.

To see a scale dependence of the effective coupling, one expands the left hand side of the arrow in the partition function(\ref{holo-partition}) as
\begin{equation}
Z=\langle \mathbb I \rangle -\frac{D^{(2)}}{2} \int d^4x\langle O^2(x) \rangle +
\frac{(D^{(2)})^2}{2^2\cdot 2!}\int d^4x d^4y \langle O^2(x)O^2(y)\rangle+ {\rm\ order\ of\ }(O^6(x)),
\end{equation}
where $\mathbb I$ is an identity operator. One may compute the operator product expansion of the $[O^2(x)O^2(y)]$ and find its coefficient of the $O^2(x)$ to see the the short distance effect of it. Such a computation gives the scale dependence of the double trace coupling as
\begin{equation}
D^{(2)}_{\rm eff}=D^{(2)}+\frac{(D^{(2)})^2}{4}2\pi^2 a\log\frac{\mu}{\Lambda}+{\rm order\ of }((D^{(2)})^3),
\end{equation}
where the $\mu$ is an energy cut-off and $\Lambda$ is a certain reference scale.

It will be very interesting to ask how gauge/gravity duality framework will realize that such a marginal double trace coupling has scale dependence.
It turns out that in the dual gravity theory, the relation(\ref{multitrace-relation}) reproduces the above result precisely. Considering a bulk free scalar theory, one may evaluate its classical solution as
\begin{equation}
\phi(r,x^\mu)=\alpha(x^\mu)r^\Delta+\beta(x^\mu)r^\Delta\log(\mu r) \rightarrow
\alpha'(x^\mu)r^\Delta+\beta'(x^\mu)r^\Delta\log(\mu' r),
\end{equation}
where $\Delta$ is the scaling dimension of the single trace operator in the dual gauge field theory and then $\Delta=\frac{d}{2}$ for the double trace coupling to become marginal. The right hand side of the arrow is the same solution with a different reference scale $\mu'$ and the coefficients $\alpha'$, $\beta'$, i.e we interpret the $\mu$ as a scale in the dual field theory. The fact that the two solution is the same gives the relations between the coefficients as
\begin{equation}
\label{RR}
\alpha(x^\mu)=\alpha'(x^\mu)+\beta'(x^\mu)\log \frac{\mu'}{\mu}, {\rm \ \ and\ \ }\beta=\beta'.
\end{equation}
On top of this, the boundary condition $\beta=D^{(2)} \alpha$ will change to $\beta'=D^{(2)\prime}  \alpha'$. We interpret this as that the coupling runs due to change of the scale in the field theory. Applying the boundary conditions to (\ref{RR}), we get
\begin{equation}
\label{dual-cft-rg}
D^{(2)\prime}=\frac{D^{(2)}}{1-D^{(2)}\log \frac{\mu'}{\mu}}.
\end{equation} 

The similar method is able to be applied to multi trace deformation cases. However, in this case, the bulk scalar field theory contains $\lambda\phi^n$ self-couplings and the bulk solution needs to be obtained up to leading order in $\lambda$. The form of multi trace deformations are given by
\begin{equation}
\sim\int d^d x \mathcal D^{(n)}\mathcal O^n(x),
\end{equation}
where $\mathcal O(x)$ is the single trace operator which couples to the boundary value of the bulk scalar field $\phi$. $\mathcal D^{(n)}$ is the multi trace coupling and we assume that $n\geq3$. For the multi trace deformation to become marginal, the conformal dimension of the single trace operator, $\mathcal O(x)$ is assigned as $\Delta=\frac{d}{n}$. When we apply the similar method as we perform above to this case, one gets
\begin{equation}
D^{(n)}(\mu)=D^{(n)}(\mu')-{\lambda}K \log\frac{\mu}{{\mu'}},
\end{equation}
where $K=1$.\footnote{We note that in \cite{Aharony:2015afa}, the $K=-\frac{1}{(-2\nu)^n}$, where $\nu=\frac{d}{2}-\Delta$. This is due to the different definitions of multi trace couplings.}

In this note, we study the massive scalar field theory with the self-interaction $\lambda \phi^n$ to explore the marginal multi trace deformations via completely different method: {\it holographic Wilsonian renormalization group(holographic Wilsoninan RG)}\cite{Heemskerk:2010hk,Faulkner:2010jy,Park:2021nyc,Park:2019pzo,Park:2018hfd,Oh:2012bx,Jatkar:2013uga,Oh:2013tsa,Oh:2015xva,Moon:2017btx}. We derive Hamilton-Jacobi type Wilsonian RG equation from the holographic model and compute the radial evolution of boundary deformations including multi trace deformations. Especially, we are interested in marginal deformation case and ask if they may show logarithmic dependence of the AdS radial variable $r$-dependence. The general solution is quite complex, so we concentrate on the leading order solutions in $\lambda$ only. 

We obtained entire solution in $r$ but interesting behavior appears near $r=0$, the UV region in the dual conformal field theory. Therefore, we expand the solution near $r=0$ and we finally obtain the scale dependence of the solutions of marginal multi trace deformations, which are given by
\begin{equation}
\label{The-crucial-expressions}
D^{(n)}(\mu)=D^{(n)}(\Lambda)-\frac{\lambda}{n}(2\pi)^{d(1-\frac{n}{2})}\log \frac{\mu}{\Lambda},
\end{equation}
where $\mu=\frac{1}{\epsilon}$, an energy scale since $\epsilon$ is inverse dimension of energy. The factor $(2\pi)^{d(1-\frac{n}{2})}$ appears since the expression(\ref{The-crucial-expressions}) is evaluated in momentum space. The  $\Lambda$ is the reference energy scale. 
\footnote{We note that in \cite{Park:2018hfd}, the authors employ holographic Wilsonian renormalization group to study classically marginal deformations. They investigate marginally, relevant or irrelevant cases in quantum level. We also note that in \cite{Grozdanov:2011aa}, the authors discuss holographic Wilsonian renormalization group for massive scalar theory with self interaction, whose form is quite general, which is given by $\sum_{n=3}^\infty \lambda_n \phi^n$. They discuss general features of the holographic Wilsonian renormalization group and its similarity with Callan-Symansik equation. However they do not give detailed discussion on the reproduction of logarithmic behaviors of margianal deformations on the boundary conformal field theory.

}

\section{Hamilton-Jacobi equation of the massive scalar theory with self-interactions in AdS space}
We start with a holographic model with massive scalar field theory with its self-interactions, which is given by
\begin{equation}
\label{model-action}
S_{bulk}=\int dr \int d^dx \ \sqrt{g} \left[\frac{1}{2} g^{\mu\nu} \partial_{\mu} \phi(r,x) \ \partial_{\nu} \phi(r,x) + \frac{1}{2} m^2 \phi^2(r,x) + \frac{\lambda}{n} \phi^n(r,x) \right]+S_B,
\end{equation}
where $S_B$ is boundary term, the background metric is 
\begin{equation}
ds^2 = g_{\mu\nu}dx^\mu dx^\nu = \frac{1}{r^2} \left( dr^2 + dx^i dx^i \right),
\end{equation}
and the index, $i$ runs from $1$ to $d$ and $m$ is the mass of the scalar field. The $i$, $j$ indices are contracted with Kronecker's $
\delta$, $\delta_{ij}$. This also means that the boundary space is Euclidean. The variable $r$ is AdS radial coordinate and $x^i$ is the boundary directional coordinate on the conformal boundary of AdS space.

\subsection{Hamilton-Jacobi equation}
In this subsection, we derive and solve Hamilton-Jacobi equation for the boundary deformation, $S_B$ by using the gravity model action(\ref{model-action}). We want to define our model in momentum space. Therefore, we perform Fourier transform with respect to the boundary coordinates, $x_i$ as
\begin{equation}
\phi(r, x_i)
\equiv
\int \frac{d^dk}{(2\pi)^{d/2}} e^{-ik_ix_i} \phi_k(r),
\end{equation}
to obtain the action(\ref{model-action}) in the boundary momentum space. The resulting action is given by
\begin{align}
S_{bulk}
&=
\int dr\sqrt{g} \int \left[\prod^{2}_{A=1}d^dk_A \right]\delta^{(d)}\left(\sum^{2}_{B=1} \vec{k_B}\right)
	\left[
			\frac{1}{2} g^{rr} \partial_r\phi_{k_1} \partial_r\phi_{k_2}
			- \frac{1}{2} (g^{ij}k_{1i}k_{2j} - m^2) \phi_{k_1} \phi_{k_2}
	\right]
	\\ \nonumber
&+ \frac{\lambda}{n} \int dr\sqrt{g} \int\left[\prod^{n}_{A=1}\frac{d^dk_A}{(2\pi)^{d/2}}
	\phi_{k_A}\right] (2\pi)^d \delta^{(d)}\left(\sum^n_{B=1}\vec{k_B}\right)
+ S_B,
\end{align}
where the indices $A$, $B$ denote momentum labels, counting number of independent momenta in the integration.
The equation of motion of this theory is given by
\begin{align}
\Pi_\phi(p) &= \sqrt{g} g^{rr} \partial_r\phi_p=\left(\frac{\delta S_B}{\delta \phi_{-p}}\right), \\
0 &=- \frac{1}{\sqrt{g}} \partial_r\Pi_\phi(p)
+ (g^{ij}p_ip_j + m^2)\phi_p
+ \frac{\lambda}{n} (2\pi)^{(1-\frac{n}{2})d}
	\int \left( \prod^{n-2}_{A=1} d^dk_A \phi_{k_A} \right) \phi_{p-\sum_{B=1}^{n-2} k_A},
\end{align}
where the first equation is the definition of canonical momentum of the field, $\phi$. The $g^{rr}=r^2$, $g^{ij}=\delta_{ij}r^2$ and $\sqrt{g}=r^{-(d+1)}$.

By using a fact that the theory will not depend on the location of the boundary, we request that
$\frac{d}{d\epsilon} S_{bulk} = 0$. This provides Hamilton-Jacobi equation for the boundary deformation $S_B$, which is given by
\begin{align}
\label{Hamilton-JacobiEQ}
\frac{\partial S_B}{\partial\epsilon}
&=
- \int_{r=\epsilon} d^dp
\left\{
		\frac{1}{2\sqrt{g}g^{rr}}\left(\frac{\delta S_B}{\delta \phi_p}\right)
									\left(\frac{\delta S_B}{\delta \phi_{-p}}\right)
		- \frac{1}{2} \sqrt{g} (g^{ij}p_ip_j + m^2)\phi_p\phi_{-p}
\right\}
\\ \nonumber
&+ \sqrt{g}\frac{\lambda}{n} \int \left[\prod^n_{A=1}\frac{d^dk_A}{(2\pi)^{d/2}}\phi_{k_A}\right] (2\pi)^d \delta^{(d)}\left(\sum^n_{B=1}\vec{k_B}\right)
\end{align}

To solve this equation, we try the following type of the solution:
\begin{align}
S_B
&=
\Lambda(\epsilon)
+ \int d^dp J(\epsilon,p)\phi_{-p}(\epsilon)
+ \sum^\infty_{m=2} \int \prod^m_{A=1} d^dp_A \phi_{p_A} \sqrt{\gamma}D^{(m)}(p_1,\ldots,p_m)
\delta^{(d)}\left(\sum^m_{B=1}\vec{p_B}\right),
\end{align}
which is a power series in $\phi$ with certain coefficients multiplied on them.
The $\gamma_{ij}$ is boundary metric, which is given by $\gamma_{ij}=\frac{1}{\epsilon^2}\delta_{ij}$ and $\sqrt \gamma=\epsilon^{-d}$.

We plug this trial solution into the Hamilton-Jacobi equation(\ref{Hamilton-JacobiEQ}) and solve the equation by assuming that the equation is an identical equation in the field $\phi$. To compare the coefficients of the $\phi^m$ in the both sides of the equation, we get the infinite number of equations which are labeled by $m$, where $m$ is a certain integer, $0<m< \infty$.

The equations for $m=1$ and $m=2$ are given by
\begin{align}
\nonumber
\partial_\epsilon\Lambda(\epsilon)
&=
- \frac{1}{2} \int d^dp \frac{1}{\sqrt{g}g^{rr}} J(\epsilon,p) J(\epsilon,-p),
\\ \nonumber
\partial_\epsilon J(\epsilon,p)
&=
-\frac{2\sqrt \gamma}{\sqrt{g}g^{rr}}J(\epsilon,p)D^{(2)}(\epsilon,p).
\end{align}


\subsection{Strategy to obtain multi trace solutions}
The forms of equations with $ m \geq 2 $ are rather complex. Therefore, we will not describe all the equations here.
\footnote{
For $ m \geq 2 $, the most general form of Hamilton-Jacobi equation is given by
\begin{align}
\label{footnoteeq}
&\delta^{(d)}\left(\sum^{m}_{B=1}\vec{p}_B\right)\partial_\epsilon[ \sqrt \gamma D^{(m)}(p_1,\ldots,p_{n'};\epsilon)]
\\ \nonumber
&=
-\int d^dp
\left.
\frac{(m+1)}{\sqrt{g}g^{rr}}J(\epsilon,-p)\sqrt \gamma D^{(m+1)}(p_1,\ldots,p_{m+1},p)\delta^{(d)}\left(\sum^{m}_{B=1}\vec{p}_B\right)
\right. \\ \nonumber
&- \frac{\gamma}{2\sqrt{g}g^{rr}}\sum^{m}_{m'=2}m'(m+2-m')
\mathcal Per
\left\{D^{(m')}\left(p_1,\ldots,p_{m'-1},-\sum^{m'-1}_{B=1}p_B;\epsilon\right)D^{(m+2-m')}\left(p_{m'},\ldots,p_{m},\sum^{m'-1}_{A=1}p_A;\epsilon\right)\right\}\delta^{(d)}\left(\sum^{m}_{B=1}\vec{p}_B\right)
\\ \nonumber
&- \frac{1}{2}\sqrt{g}(-g^{ij}p_{1i}p_{2j}+m^2)\delta^{(d)}(\vec{p_1}+\vec{p_2})\delta_{m,2}
+ \left. \sqrt{g}\ \frac{\lambda}{n}\delta_{n,m}(2\pi)^{d(1-\frac{n}{2})}\delta^{(d)}\left(\sum^m_{B=1}\vec{p}_B\right)
\right.
\end{align}
}
Rather than this, we will impose restriction on the equations to simplify them. In fact, our purpose is to obtain radial evolution of multi-trace operators. Among them, we are especially interested in the parts of the solutions which are leading order in the self-interaction coupling $\lambda$ of the scalar field in the gravity model. 

To see the leading effects from the scalar field self-interaction, we suppose that
\begin{itemize}
\item The first restriction: all the $J_k=0$.
\item The second restriction: all the $D^{(m)}(k_1,...,k_m;\epsilon)=0$ for $2<m<n$.
\end{itemize}

The first restriction means that all the single trace operator deformation vanishes. 
The dual conformal field theory has no non-trivial deformations from the single trace operators at all and the theory may be conformal. However, even though this happens, we might have non-trivial evolution of multi-trace deformation as the energy scale changes which will break the conformal symmetry.

In fact, the equations obtained by comparison of the coefficients of $\phi^m$ in the both sides of equation(\ref{Hamilton-JacobiEQ})(we call it $m$-th equation), where $2<m<n$, are given by the following form,
\begin{eqnarray}
\label{m-theq}
\partial_\epsilon [\sqrt \gamma D^{(m)}(p_1,...,p_{m};\epsilon)]&=&2m\gamma\mathcal Per\{D^{(2)}(p_1,-p_1;\epsilon)D^{(m)}(p_1,...,p_{m};\epsilon)\}\\ \nonumber
&+&{\rm bilinears\ of\ }D^{(n')}, {\rm \ for\ }2<n'<m,
\end{eqnarray}
where $\mathcal Per\{\}$ means that 
\begin{equation}
\mathcal Per\{M(p_1,...,p_n)\}=\frac{1}{n!}(M(p_1,...,p_n)+{\rm all\ other\ permutations\ in\ momenta})
\end{equation}
where $M(p_1,...,p_n)$ is a momentum dependent function.
The bilinears of $D^{(n')}$ denotes that
\begin{equation}
{\rm the\ bilinears\ of\ }D^{(n')}\sim\sum_{n'=3}^{m-1}\alpha_{n'}\gamma D^{(n')}D^{(m+2-n')},
\end{equation}
with a certain real coefficient $\alpha_{n'}$. With careful observation on the equation(\ref{m-theq}), one can argue that if all the $D^{(n')}(k_1,...,k_{n'};\epsilon)=0$, then one can get $D^{(m)}(k_1,...,k_{m};\epsilon)=0$ solution for $2<n'<m$.
We note that the most general solution of $D^{(n')}(k_1,...,k_{n'};\epsilon)\neq0$. However, these solutions has no $\lambda$ dependence at all. Since our purpose is to get leading order contribution of the coupling $\lambda$ to the solutions, these solutions are not useful to us.

To get solutions of multi trace couplings in leading order in $\lambda$, we will look at the solutions which satisfy the first and the second restrictions. To get such a solution, we look at only two equations among them, which are the equations for $m=2$ and $m=n$.

Let us examine $m=2$ case first. Since we set $J_k=0$, the equation becomes
\begin{equation}
\partial_\epsilon[ \sqrt\gamma D^{(2)}(p,-p)]
=
- \frac{2}{\sqrt{g}g^{rr}}\left[\sqrt\gamma D^{(2)}(p,-p)\right]^2
- \frac{1}{2}\sqrt{g}(g^{ij}p_ip_j + m^2),
\end{equation}
where $D^{(2)}(p,-p)$ is called double trace coupling. The solution is well known\cite{Oh:2015xva}, which is given by
\begin{align}
\sqrt \gamma D^{(2)}(p,-p) = \frac{1}{2} \frac{\Pi_{\phi}(p,\epsilon)}{\phi_p(\epsilon)},
\end{align}
where $\Pi_\phi(p)$ and $\phi_p$ satisfy the following equations:
\begin{align}
\label{canonical-sol}
\Pi_\phi(p,\epsilon)
&=
\sqrt{g(\epsilon)}g^{rr}(\epsilon)\partial_\epsilon \phi_p(\epsilon)
\\
\label{bulk-equations-sol}
0
&=
- \frac{1}{\sqrt{g(\epsilon)}}\partial_\epsilon\Pi_{\phi}(p,\epsilon) + (g^{ij}p_ip_j+m^2)\phi_p(\epsilon),
\end{align}
where $g_{\mu\nu}(\epsilon)$ is the bulk metric tensor evaluated at $r=\epsilon$. 
It is also known that the solutions of the equations(\ref{canonical-sol}) and (\ref{bulk-equations-sol}) are given by
\begin{align}
\label{sol-no-momentum}
\phi (\epsilon) = \alpha \epsilon^{\frac{d}{2}-\nu} + \beta \epsilon^{\frac{d}{2}+\nu}
\end{align}
when the boundary directional momentum vanishes. $\alpha$ and $\beta$ are arbitrary constants. When we have non-zero boundary directional momenta, the solution becomes
\begin{align}
\label{KI-solution}
\phi_p(\epsilon) = \alpha(p)\epsilon^{\frac{d}{2}}K_\nu (|p|\epsilon) + \beta(p)\epsilon^{\frac{d}{2}} I_\nu (|p|\epsilon),
\end{align}
where $\alpha(p)$ and $\beta(p)$ are arbitrary boundary momentum $p$-dependent functions. The parameter $\nu = \frac{1}{2}\sqrt{d^2 + 4m^2}$.

Now we consider $m=n\geq 2$ case and we also assume that $J_p=0$ too. The equation in this case is given by
\begin{align}
\nonumber
\partial_\epsilon[\sqrt\gamma  D^{(n)}(p_1,\ldots,p_n)]
&=
- \frac{\gamma}{2\sqrt{g}g^{rr}}\sum^n_{m=2}m(n+2-m)D^{(m)}(p_1,\ldots,p_{m-1},-\sum^{m-1}_{B=1}p_B) \\ 
&\times
D^{(n+2-m)}(p_m,\ldots,p_n,\sum^{m-1}_{A=1}p_A) + \sqrt{g}\frac{\lambda}{n}(2\pi)^{d(1-\frac{n}{2})}.
\end{align}
Since we restrict ourselves in the case that 
$D^{(m)} = 0$ for $n>m>2$, this equation is simplified to
\begin{align}
\nonumber
\partial_\epsilon [\sqrt\gamma D^{(n)}(p_1,\ldots,p_n)]
&=
- \frac{2n\gamma}{\sqrt{g}g^{rr}} \mathcal Per\{D^{(2)}(p_1,-p_1)D^{(n)}(p_1,p_2,\ldots,p_n)\}
+ \sqrt{g}\frac{\lambda}{n}(2\pi)^{d(1-\frac{n}{2})}
\\
&=
\frac{-2\gamma}{\sqrt{g}g^{rr}}\left(\sum^n_{A=1}D^{(2)}(p_A,-p_A)\right)D^{(n)}(p_1,\ldots,p_n)
+ \sqrt{g}\frac{\lambda}{n}(2\pi)^{d(1-\frac{n}{2})}
\label{n-th trace equ}
\end{align}

Since the solution of the double trace coupling is given by $\sqrt \gamma D^{(2)}(p_A,-p_A)=\frac{1}{2}\frac{\Pi_\phi({p_A})}{\phi_{p_A}} = \frac{1}{2}\sqrt{g(\epsilon)}g^{rr}(\epsilon)\frac{\partial_\epsilon\phi_p(\epsilon)}{\phi_p(\epsilon)}$, we plug this solution into (\ref{n-th trace equ}). Then, we have
\begin{equation}
\label{the final equation}
\partial_\epsilon [\sqrt\gamma D^{(n)}(p_1,\ldots,p_n)]
=
- \sqrt\gamma \left( \sum^n_{A=1}\partial_r\log\phi_{p_A}(\epsilon) \right)D^{(n)}(p_1,\ldots,p_n)
+ \sqrt{g}\frac{\lambda}{n}(2\pi)^{d(1-\frac{n}{2})}.
\end{equation}



The form of the solution of (\ref{the final equation})  is given by 
\begin{equation}
\label{multi-trace-solution}
D^{(n)}(p_1,...,p_n;\epsilon)=D^{(n)}_H(p_1,\ldots,p_n;\epsilon) +D^{(n)}_\lambda(p_1,...,p_n;\epsilon),
\end{equation}
where the solution is sum of two different types of the solutions of this equation. One is homogeneous solution meaning the solution when the coupling $\lambda$ is turned off and another is in-homogeneous solution which is the solution being proportional to the coupling $\lambda$.

The homogeneous solution is given by
\begin{equation}
D^{(n)}_H(p_1,\ldots,p_n;\epsilon) = \frac{C^{(n)}}{\sqrt \gamma}\prod^n_{A=1}[\phi_{p_A}(\epsilon)]^{-1},
\end{equation}
and the in-homogeneous solution of the equation(\ref{the final equation}) is given by
\begin{align}
\label{inhomogeneous-sol}
D^{(n)}_\lambda(p_1,...,p_n;\epsilon)
&=\frac{\lambda}{n}(2\pi)^{d(1-\frac{n}{2})}
\left[\frac{\int^\epsilon_{\Lambda^{-1}} d\epsilon ' \prod^n_{A=1}\left[\phi_{p_A}(\epsilon ')\right] \sqrt{g(\epsilon')}}{\prod^n_{B=1}\left[\phi_{p_B}(\epsilon)\right]\sqrt{\gamma(\epsilon)}}\right],
\end{align}
where the $\Lambda$ is a constant with dimension of energy to  set the initial value of the variable $\epsilon$ and $C^{(n)}$ is a real constant.

\section{Renomalization group evolution of marginal multi trace deformation}

In this section, we consider a case that multi trace couplings given in the boundary deformation $S_B$ is classically marginal. The multi trace deformation has a form of
\begin{equation}
{\rm the\ multi\ trace\ deformation}=\int d^d x \mathcal D^{(n)}\mathcal O^n(x),
\end{equation}
where $\mathcal O(x)$ is the single trace operator which couples to the boundary value of the bulk scalar field $\phi$. $\mathcal D^{(n)}$ is the multi trace coupling and we assume that $n\geq3$. 

Now we take the scaling dimension of the operator $\mathcal O(x)$ to be  $\Delta=\frac{d}{n}$.
With this, let us evaluate the evolution of the multi trace solution(\ref{multi-trace-solution}) explicitly by using the solution $\phi_p(\epsilon)$ which is given in (\ref{KI-solution}) near the UV-fixed point(or one can use the solution (\ref{sol-no-momentum}) for zero boundary momentum case). The UV-fixed point means that near $\epsilon=0$.
Again we want to evaluate the evolution of the multi trace couplings near $\epsilon=0$, we suppose that the $\Lambda^{-1}<\epsilon << 1$ in the expression (\ref{inhomogeneous-sol}).

Near $\epsilon=0$, the solution of $\phi_p$ is expanded as a power series in small $\epsilon$, whose leading behavior is given by
\begin{equation}
\label{nearepsilonzero}
\phi_p(\epsilon)=\mathcal A(p)\epsilon^\Delta + \mathcal B(p)\epsilon^{d-\Delta},
\end{equation}
where $\mathcal A(p)$ and $\mathcal B(p)$ are coefficients of normalizable and non-normalizable modes of the solution respectively.

Now examine the homogeneous solution.  Recall we choose that $\Delta=\frac{d}{n}$ for the $D^{(n)}(p_1,\ldots,p_n)$ to become classically marginal. Near $\epsilon=0$, the solution is given by
\begin{eqnarray}
D^{(n)}_H(p_1,\ldots,p_n;\epsilon) &=&\frac{ C^{(n)}}{\sqrt{\gamma}}\prod^n_{A=1}[\phi_{p_A}(\epsilon)]^{-1},\\ \nonumber
&\rightarrow& C^{(n)}\left[\prod_{A=1}^n \mathcal A^{-1}(p_A)\right](1+O(\epsilon^{d(1-\frac{2}{n})}))
\end{eqnarray}
As one observes, the multi trace coupling is nearly a constant up to leading order in $\epsilon$. 
Once we define a new constant as 
\begin{equation}
D^{(n)}_H(p_1,\ldots,p_n;\Lambda)\equiv C^{(n)}\prod_{A=1}^n \mathcal A^{-1}(p_A),
\end{equation}
then the near $\epsilon=0$ solution becomes
\begin{equation}
D^{(n)}_H(p_1,\ldots,p_n;\epsilon)=D^{(n)}_H(p_1,\ldots,p_n;\Lambda)+O(\epsilon^{d(1-\frac{2}{n})}),
\end{equation}
up to leading order in $\epsilon$, which is almost a constant meaning that there is no extra logarithmic behavior.

Next, let us look at the in-homogeneous solution. Using the leading behaviors of the solution(\ref{nearepsilonzero}) near $\epsilon=0$, one can evaluate the following factors in (\ref{inhomogeneous-sol}):
\begin{eqnarray}
\sqrt{\gamma}(\epsilon)\prod^n_{A=1}[\phi_{p_A}(\epsilon)] \rightarrow \prod_{A=1}^n \mathcal A(p_A)(1+O(\epsilon^{d(1-\frac{2}{n})})), \\
\sqrt{g}(\epsilon)\prod^n_{A=1}[\phi_{p_A}(\epsilon)] \rightarrow \prod_{A=1}^n \mathcal A(p_A)(\epsilon^{-1}+O(\epsilon^{d(1-\frac{2}{n}-\frac{1}{d})})).
\end{eqnarray}
With such expressions, we get the near $\epsilon=0$ behavior of the in-homogeneous solution as
\begin{equation}
D^{(n)}_\lambda(p_1,...,p_n;\epsilon)\rightarrow \frac{\lambda}{n}(2\pi)^{d(1-\frac{n}{2})}\log \frac{\epsilon}{\Lambda^{-1}}+O(\epsilon^{d(1-\frac{2}{n})}).
\end{equation}

Finally we can conclude that the total solution for the multi trace deformation near $\epsilon=0$ is given by
\begin{eqnarray}
\label{conclusive-formula}
D^{(n)}(p_1,...,p_n;\epsilon)&=&D^{(n)}(p_1,...,p_n;\Lambda)+\frac{\lambda}{n}(2\pi)^{d(1-\frac{n}{2})}\log \frac{\epsilon}{\Lambda^{-1}}, \\ \nonumber
&=&D^{(n)}(p_1,...,p_n;\Lambda)-\frac{\lambda}{n}(2\pi)^{d(1-\frac{n}{2})}\log \frac{\mu}{\Lambda}
\end{eqnarray}
where we define $\mu\equiv\frac{1}{\epsilon}$ as an energy scale since $\epsilon$ is inverse of energy scale.
We interpret that $D^{(n)}(p_1,...,p_n;\Lambda)$ is the multi trace coupling at energy scale of $\Lambda$ and $D^{(n)}(p_1,...,p_n;\mu)$ is the one at different energy scale, $\mu$. The two couplings are as different as the logarithmic term appeared in (\ref{conclusive-formula}).

\section*{Acknowledgment}
J.H.O thank his W.J. and Y.J. This work was supported by the National Research
Foundation of Korea(NRF) grant funded by the Korea government(MSIP) (No.2016R1C1B1010107).

\end{document}